\documentclass[12pt, draftclsnofoot, onecolumn]{IEEEtran}
\usepackage{stmaryrd}
\usepackage{amsmath}
\usepackage{amssymb}
\usepackage{psfrag}
\usepackage{graphicx}
\usepackage{epstopdf}
\usepackage{todonotes}
\usepackage{cite}
\usepackage{algorithmic}
\usepackage{algorithm}
\usepackage{svg}
\usepackage{multirow}

\newcommand{\num}{{}}

\begin{document}
\title{RCNet: Incorporating Structural Information into Deep RNN for MIMO-OFDM Symbol Detection with Limited Training}

\author{Zhou Zhou, Lingjia Liu, Shashank Jere, Jianzhong (Charlie) Zhang, and Yang Yi
\thanks{Z. Zhou, L. Liu, S. Jere, and Y. Yi are with Bradley Department of Electrical and Computer Engineering, Virginia Tech, USA. J. Zhang is with Standards and Mobility Innovation Lab., Samsung Research America, USA. The work of Z. Zhou, L. Liu, S. Jere, and Y. Yi are supported in part by U.S. National Science Foundation under grants: CCF-1937487 and ECCS-1811497. Part of the work will be presented in~\cite{zhou2020AAAI}} \thanks{The corresponding author is L. Liu (ljliu@ieee.org).}
}
\maketitle
\begin{abstract}
In this paper, we investigate learning-based MIMO-OFDM symbol detection strategies focusing on a special recurrent neural network (RNN) -- reservoir computing (RC). 
We first introduce the Time-Frequency RC to take advantage of the structural information inherent in OFDM signals.
Using the time domain RC and the time-frequency RC as the building blocks, we provide two extensions of the shallow RC to RCNet: 1) Stacking multiple time domain RCs; 2) Stacking multiple time-frequency RCs into a deep structure. 
The combination of RNN dynamics, the time-frequency structure of MIMO-OFDM signals, and the deep network enables RCNet to handle the interference and nonlinear distortion of MIMO-OFDM signals to outperform existing methods. 
Unlike most existing NN-based detection strategies, RCNet is also shown to provide a good generalization performance even with a limited training set (i.e, similar amount of reference signals/training as standard model-based approaches). 
Numerical experiments demonstrate that the introduced RCNet can offer a faster learning convergence and as much as $20\%$ gain in bit error rate over a shallow RC structure by compensating for the nonlinear distortion of the MIMO-OFDM signal, such as due to power amplifier compression in the transmitter or due to finite quantization resolution in the receiver.
\end{abstract}

\section{Introduction}
\label{Sec:Intro}



Artificial-intelligence (AI) enabled cellular network is envisioned as the critical path for beyond 5G and 6G networks~\cite{DBLP:journals/corr/abs-1907-07862,AI_IoT_Hao}.
Among the various potential fields of communication systems where AI and its associated machine learning tools can contribute, symbol detection is a very important one in the physical layer.
To be specific, symbol detection constitutes a key module within the signal processing chain of modern communication receivers. 
Assuming the availability of receiver channel state information (CSI), the optimal model-based strategy is to apply the maximum likelihood detector. 
However, the performance of model-based strategies is sensitive to model inaccuracies and CSI estimation errors. 
On the other hand, learning-based approaches can provide robust performance without relying on detailed underlying channel models.
For example, the method introduced in \cite{samuel2019learning} shows that an unfolding neural network (NN) from the projected gradient descent algorithm can be trained to achieve state-of-the-art performance for MIMO symbol detection tasks. 
In this paper, we focus on the problem of symbol detection in MIMO-OFDM which is the major radio access technology for 4G/5G systems~\cite{LTE,5GNR}.
In current 4G/5G systems, symbol detection methods are based on modeling the underlying wireless link and applying associated model-based signal processing techniques~\cite{yang2015fifty}. 
However, in the presence of non-linearities, either due to the underlying wireless channels (e.g. mmWave and Terahertz channels for 5G-Beyond) or device components (e.g. power amplifier), it becomes extremely difficult to analytically model such behaviors in a tractable and accurate manner. 
Alternatively, recent work in \cite{zhao2018deep, chen2019efficient, Samuel2017, jiang2018artificial} demonstrated the effectiveness of using NNs for symbol detection under unknown environments. 
In the same line of thinking, we consider exploiting the dynamic behaviors of recurrent neural networks (RNNs) for the task of MIMO-OFDM symbol detection. 
The main motivation of adopting RNNs instead of other NN architectures is based on the fact that under fairly mild and general assumptions, RNNs are universal approximations of dynamic systems~\cite{FUNAHASHI1993801}. 
This is extremely important for wireless systems which are highly dynamic over time and frequency. 
On the other hand, to realize the full potential of RNNs, especially deep RNNs, new research challenges and issues need to be addressed for MIMO-OFDM symbol detection:
\begin{itemize}
\item \textbf{Challenge 1}: From a system design perspective, training a NN-based symbol detector using over-the-air feedback --- to update layer weights of the underlying NN based on the back-propagation algorithm --- is likely prohibitively expensive in terms of control overhead\footnote{For example, in 3GPP LTE/LTE-Advanced systems, the reference signal overhead is specified and is usually fixed for different MIMO configurations~\cite{LTE_standards}: The training set (demodulation reference signals) for SISO-OFDM is around $5\%$ of all the resource elements. 
On the other hand, for a $2 \times 2$ MIMO-OFDM system, the overhead for reference signals is around $10\%$.
In 5G, more flexible reference signal design is designed to reduce the reference signal overhead~\cite{5GNR2}}.
Therefore, over-fitting could happen when the selected NN model is too complicated. 
\item \textbf{Challenge 2}: NNs, especially RNNs, are mainly designed to process time-domain data. Since data is transmitted in both time and frequency domains in contemporary cellular systems, it is critical to combine RNN with domain knowledge in an organic way to offer reliable and robust performance gains over current communication strategies. 
\item \textbf{Challenge 3}: The underlying wireless environment changes dynamically over time and frequency. This is especially true for 5G and future 6G systems which mainly use mmWave and Terahertz channels. Accordingly, the underlying NN model needs to be sophisticated enough to capture the time-frequency variation of the channel. Otherwise, under-fitting can result in poor symbol detection performance. 
\end{itemize}
\subsection{Our Contributions}
To address these challenges, our approach is based on reservoir computing (RC)~\cite{lukovsevivcius2009reservoir} which is a special category of RNNs. 
RC is capable of avoiding the issues of gradient vanishing and exploding which occur during training of conventional RNNs using back-propagation through time (BPTT)~\cite{pascanu2013difficulty}. 
Furthermore, the training of RC is only conducted on the output layer while the hidden layers and the input layers are fixed according to a certain distribution.
In this way, the amount of training needed for MIMO-OFDM symbol detection can be significantly reduced making it an \textbf{\emph{operationally feasible}} solution to address \emph{Challenge 1}. 
This benefit can be clearly seen in Section~\ref{subsec:Comparison} where show a quantitative comparison on the training overhead for various learning-based strategies.
The RC architecture makes applying NN techniques in the physical layer of cellular networks possible and feasible.
Furthermore, it is shown in~\cite{RubayetRC} that the RC-based symbol detection can significantly improve the underlying energy-efficiency of the system.

In this work, rather than directly applying the shallow RC structure, we provide an attempt to address \emph{Challenge 2} and \emph{Challenge 3} by introducing \textit{RCNet} through the following extensions to facilitate \textbf{\emph{deep}} RC-based symbol detection methods to further improve detection performance:
\begin{itemize}
\item Extend the output layer of a shallow RC structure to a multiple-layer network promoting the joint time-frequency processing neuron states. This method can effectively resolve \emph{Challenge 2} by incorporating the structural information (time-frequency structure) of MIMO-OFDM signals into the output layer design of RCNet.
\item Stack shallow RCs together into a ``deep'' RC to improve the processing capability of RCNet. In this way, \emph{Challenge 3} can be addressed owing to the boosting mechanism of NNs.
\end{itemize}

The first extension on deepening the output layer is achieved by replacing the original single layer output of the shallow RC to a three-layer structure: a time-domain layer, a Fourier transform layer, and a frequency-domain layer, namely the ``time-frequency RC''. 
The time-domain layer attempts to reconstruct the transmitted time-domain signal. 
The Fourier transform layer is used to transform the time-domain signal to the frequency-domain. 
The frequency-domain layer attempts to extract frequency-domain features to further improve the detection performance. 
The second extension is achieved by concatenating multiple ``time-frequency RCs'' sequentially. 
Note that the output weights of each RC layer are trained in a consecutive fashion.

Through extensive experiments, we show that this deep structure of RCNet demonstrates very appealing and robust performance when non-linear effects exist in the end-to-end wireless system. 
In other words, it outperforms conventional MIMO symbol detection strategies under channel non-linearities caused by the power amplifier (PA) at the transmitter (Tx) and/or the quantization error due to the low resolution of analog-to-digital converters (ADCs) at the receiver (Rx).
The results suggest that the introduced RCNet MIMO-OFDM symbol detection can be a very promising enabling technology for 5G-Beyond and 6G cellular systems where high-frequency spectrum and low-resolution ADCs will be frequently used.
\subsection{Related Work}
\subsubsection{Deep RNN}
Deep neural networks (DNN) can extract sophisticated features thereby providing improved classification performance over shallow NNs~\cite{hinton2012deep}. 
Hierarchical RNNs are shown to be capable of learning long-term dependencies of signals~\cite{el1996hierarchical}. 
In \cite{Grave2013}, the deep long short-term memory network was introduced by consecutively stacking the hidden layers of multiple RNNs. 
This deep structure is shown to significantly improve the performance in the task of speech recognition. 
The methods of extending a shallow RNN to deep RNNs have been summarized in~\cite{pascanu2013construct}: multi-layer RNNs can be constructed by increasing input layers, hidden layers and output layers, as well as stacking multiple shallow RNNs into a deep form. 
Concatenating echo state networks (ESNs) into a chain by learning readout layers connecting to each ESN layer is introduced in \cite{jaeger2007discovering}. 
The follow up work in \cite{gallicchio2017echo, gallicchio2017deep}
extended this structure into a deep version which is demonstrated to be able to achieve a higher memory compared to the shallow one. 
However, these extensions are very general without considering specific domain knowledge and structural information for targeted applications.
\subsubsection{NN based Symbol Detection}
Based on the convolution feature of wireless channels, convolutional neural networks (CNNs) is utilized for OFDM symbol detection~\cite{zhao2018deep}. 
The work in \cite{Samuel2017} combines the CSI and received symbols together as the input of DNNs for MIMO symbol detection. 
In \cite{chen2019efficient}, it is shown that the NN-based MIMO symbol detection can be generalized under imperfect CSI. 
Two frameworks, data-driven and model-driven NNs, are introduced in~\cite{jiang2018artificial} for the design of an OFDM receiver. However, all of these methods are based on general NNs requiring a large amount of training set.
Therefore, it is almost impossible to adopt any of these methods in practical cellular networks.

On the other hand, our previous work~\cite{mosleh2017brain, zhou2019} showed that shallow RCs can achieve good performance for MIMO-OFDM symbol detection even with very limited training set. 
In~\cite{mosleh2017brain}, an ESN, a special case of RC, is introduced as an OFDM symbol detector without relying on obtaining explicit CSI. This scheme is evaluated under relevant scenarios for cellular networks where the training/reference signal overhead is comparable to that of current cellular networks. 
In our follow up work,  windowed ESN (WESN) is introduced by adding a sliding window to the input of ESN to enhance the short-term memory (STM)~\cite{zhou2019}. 
Experimental results show that WESN can provide good performance over standard ESNs using the training/reference signal set adopted in 4G LTE-Advanced~\cite{LTE}. 
Furthermore, the ESN-based symbol detectors can effectively compensate for the distortion caused by non-linear components of the wireless transmission.

The organization of this paper is as follows. In Sec. \ref{MIMO_OFDM_data}, we briefly describe the transceiver architecture. In Sec. \ref{Time_Frequency_RC} and \ref{Layered_RC0}, we introduce the two extensions incorporated in our proposed RCNet structure and its associated learning algorithms. Sec. \ref{Numerical_Experiments} evaluates the performance of RCNet as opposed to existing symbol detection strategies for MIMO-OFDM systems. 
The conclusion and future work is contained in Sec. \ref{Conclusion}.

\section{OFDM Transceiver Architecture and Structrual Information}
\label{MIMO_OFDM_data}

In this section, we briefly introduce the transceiver architecture of a MIMO-OFDM system and illustrate the structural information we are going to utilize for the design of RCNet.  

\begin{figure}[!h]
	\centering
	\includegraphics{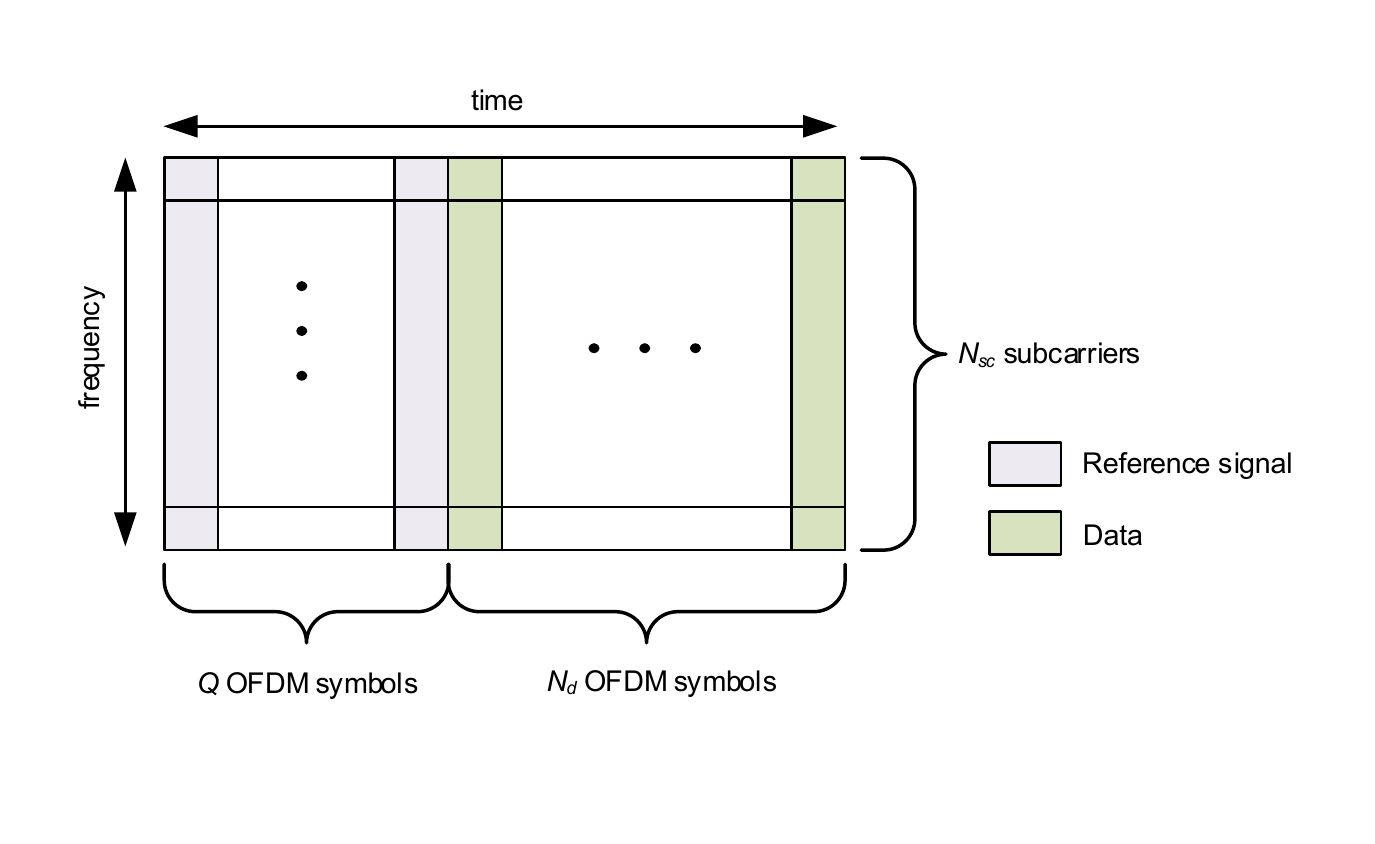}
	\caption{OFDM Resource Grid}
	\label{OFDM_resource_grid}
\end{figure}

The resource grid of the OFDM can be seen most clearly in Fig.~\ref{OFDM_resource_grid}.
The total system bandwidth is divided into $N_{sc}$ sub-carriers.
In each sub-frame, the first $Q$ OFDM symbols are the reference signals (the training set) and the rest $N_{d}$ OFDM symbols are used to carry data (the testing set). 
Therefore, the reference signal overhead (training overhead) is defined as $\eta = Q / (Q+N_{d})$.
In 3GPP LTE/LTE-Advanced and 5G systems, $Q+N_{d} = 14$ for normal sub-frames and the overhead is typically below $20\%$ to improve spectral-efficiency~\cite{LTE_standards,5GNR2}.
For a single MIMO-OFDM symbol, we denote the modulation symbol as $\{{\boldsymbol z}(n)\}_{n=0}^{N_{sc}-1}$, where ${\boldsymbol z}(n) \in {\mathbb C}^{N_t \times 1}$ represents the modulation symbols at the $n$th sub-carrier of the underlying MIMO-OFDM symbol in the {\textbf {frequency domain}}; $N_t$ stands for the number of transmitting antennas in the system. 
Each element of ${\boldsymbol z}(n)$ is modulated by quadrature amplitude modulation (QAM). A single MIMO-OFDM symbol in the frequency domain also can be lumped as a matrix,
\begin{align}
{\boldsymbol Z} \triangleq [{\boldsymbol z}(0), {\boldsymbol z}(1), \cdots, {\boldsymbol z}(N_{sc}-1)]^T.
\end{align}
After the inverse fast Fourier transform (IFFT) operation and addition of the cyclic prefix (CP), we can obtain the \textbf{time domain} MIMO-OFDM symbol, denoted as $\boldsymbol {\tilde X}$. At the transmitter, the MIMO-OFDM symbol is distorted by a non-linear activation function $f(\cdot)$ due to the inherent non-linearity of transmitter-side radio circuits, such as the PA~\cite{joung2014survey}. At the receiver, a single MIMO-OFDM symbol is expressed as 
\begin{align}
{\boldsymbol X}= q(h(f({\boldsymbol {\tilde X}})) + {\boldsymbol N}),
\end{align}
where $\boldsymbol X \triangleq [{\boldsymbol x}(0), {\boldsymbol x}(1), \cdots, {\boldsymbol x}(N_{sc}+N_{cp}-1)]^T \in {\mathbb C}^{(N_{sc}+N_{cp}) \times N_r}$; ${\boldsymbol N}$ is the additive noise; $h(\cdot)$ stands for the multi-path channel, such as the 3GPP spatial channel model (SCM) \cite{FD_MIMO_model} and $q(\cdot)$ represents the non-linearity at the receiver.

For RC-based MIMO-OFDM symbol detection, the objective is to recover the {\textbf{frequency domain}} modulation symbols $\{{\boldsymbol z}(n)\}_{n=0}^{N_{sc}-1}$. 
On the other hand, the input of the RC is the {\textbf{time domain}} samples $\{{\boldsymbol x}(t)\}_{t=0}^{N_{cp}+N_{sc}-1}$. 
This structural information needs to be explored in the design of RCNet to further improve the detection performance on top of the existing shallow RC-based symbol detection.
In the supervised learning framework, the training set is defined as $\{d_q\}_{q = 0}^{Q-1}$:
\begin{equation}
\begin{aligned}
d_q &\triangleq \Big(\{{\boldsymbol x}_q(t)\}_{t=0}^{N_{cp}+N_{sc}-1},\{{\boldsymbol z}_q(n)\}_{n=0 }^{N_{sc}-1}\Big)\\
& \cong \Big(\{{\boldsymbol x}_q(t)\}_{t=0}^{N_{cp}+N_{sc}-1},\{{\boldsymbol {\tilde x}}_q(t)\}_{t=0 }^{N_{cp}+N_{sc}-1}\Big)\\
&\cong \big( {\boldsymbol X}_q, {\boldsymbol Z}_q\big) \cong \big( {\boldsymbol X}_q, {\boldsymbol {\tilde X}}_q\big)
\end{aligned};
\label{Training_batch}
\end{equation}
$\cong$ represents equivalently defined as; The subscript $q$ stands for the $q$th MIMO-OFDM symbol, i.e, the training set has $Q$ batches in total. 
The notations are summarized in Table~\ref{notations}.
\begin{table}[!h]
\caption{Notations}
\centering
\begin{tabular}{|c|c|}
\hline
\textbf{Symbols}   & \textbf{Definitions}                     \\
\hline
$N_r$    & Number of receiver antennas       \\
\hline
$N_t$    & Number of transmitter antennas    \\
\hline
$N_{sc}$    & Number of sub-carriers             \\
\hline
$N_{cp}$ & Length of Cyclic Prefix (CP)          \\
\hline
$Q$      &  Number of MIMO-OFDM symbols in the training set (Number of batches in the training set) \\
\hline
$N_{d}$      &  Number of MIMO-OFDM symbols carrying data (testing set) \\
\hline
$\eta$       &  Reference signal (training) overhead \\
\hline
$\boldsymbol X$    & One MIMO-OFDM symbol at Rx in time domain (One batch of training input) \\
\hline
$\boldsymbol {\tilde X}$    & One MIMO-OFDM symbol at Tx in time domain (One batch of training target in time domain) \\
\hline
$\boldsymbol Z$    & One MIMO-OFDM symbol at Tx in frequency domain\\
&(One batch of training target in frequency domain)\\
\hline
${\boldsymbol x}(t)$ & The $t$th sample of one MIMO-OFDM symbol at Rx in the time domain \\
\hline
${\boldsymbol {\tilde x}}(t)$ & The $t$th sample of one MIMO-OFDM symbol at Tx in the time domain \\
\hline
${\boldsymbol z}(n)$ & The $n$th modulation symbol of one MIMO-OFDM symbol at Tx in the frequency domain \\
\hline
\end{tabular}
\title{Table of Notations}
\label{notations}
\end{table}


\section{Time-Frequency RC -- Incorporating Structural Information}
\label{Time_Frequency_RC}
To incorporate the time-frequency structural information inherent in the OFDM signal structure, we will introduce the new concept of ``time-frequency RC'' on top of the shallow RC in this section. 
For differentiation, the shallow RC is referred as the ``time domain RC'' in this paper. 
To provide a systematic view of the RC-based design and to better articulate how the structural information is incorporated, we will first briefly discuss the ``time domain RC'' used in our previous work~\cite{mosleh2017brain,zhou2019} before describing the ``time-frequency RC''.

\subsection{Time Domain RC -- The Shallow RC}
\label{subsec:TRC}
\begin{figure}[!h]
\centering
\includegraphics[width=0.75\linewidth, height = 0.375\linewidth]{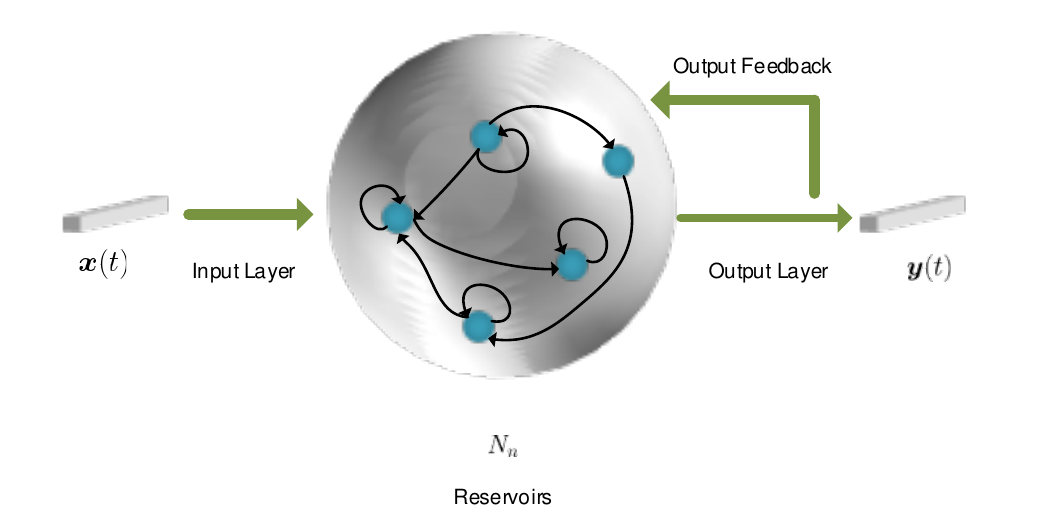}
\caption{RCNet: Time Domain RC - Echo State Network (ESN) Realization}
\label{ESN}
\end{figure}
One realization of the time domain RC is illustrated in Fig. \ref{ESN}, namely an ESN~\cite{jaeger2001echo}. 
In the figure, the collection of neurons is denominated as a \textit{reservoir}. 
The ESN drives the input signal into a high dimensional dynamic response through a \textbf{fixed} random projection, where the response signal is represented by the trajectory of hidden neuron states. 
Meanwhile, non-linear activation functions are applied to the neuron states transition. 
The neurons in the reservoir are sparsely connected with fixed weights to satisfy certain distributions under which the response signals are asymptotically uncorrelated to the initial neuron states \cite{jaeger2001echo}. 
One justification for using fixed hidden states transition is from an experimental fact that the dominant changes of an RNN's weights during training happen at the output layer~\cite{schiller2005analyzing}. 
Finally, the desired outputs are obtained by learning a combination of the non-linear response signals.

Given the training set $\{d_q\}_{q=0}^{Q-1}$, the states of the reservoir generated by the $q$th batch are
\begin{align}
\label{Dynamic_Eq}
{\boldsymbol s}_q(t+1) &= f({\boldsymbol s}_q(t){\boldsymbol W}_s+{\boldsymbol x}_q(t){\boldsymbol W}_{in}+{\boldsymbol {\tilde x}_q}(t){\boldsymbol W}_{fb}+{\boldsymbol n}(t))
\end{align}
where ${\boldsymbol s}_q(t) \in {\mathbb C}^{N_n\times 1}$ represents the neurons state vector and $N_n$ denotes the number of neurons in the reservoir; $f$ is the activation function; ${\boldsymbol W}_s$ denotes the state transition weights; ${\boldsymbol W}_{in}$ denotes the weights of input layer; ${\boldsymbol n}(t)$ is an optional noise regularization term; ${\boldsymbol W}_{fb}$ represents the weights on the feedback path which can be removed when teacher forcing is not required \cite{jaeger2001echo}. Correspondingly, the output signal of the time domain RC can be written as
\begin{align}
    {\boldsymbol y}_q(t) = {\boldsymbol s}_q(t){\boldsymbol W}_{tout}
    \label{time_output}
\end{align}
where ${\boldsymbol W}_{tout}$ represents the readout layer.

To drive ${\boldsymbol y}_q(t)$ to the desired time domain MIMO-OFDM symbol, we can minimize the $l_2$ norm distance between $\{{\boldsymbol y}_q(t)\}_{q=0}^{Q-1}$ and $\{{\boldsymbol {\tilde x}}_q(t)\}_{q=0}^{Q-1}$ through
\begin{align}
\label{time_objective}
\min_{{\boldsymbol W}_{tout}} \sum_{q=0}^{Q-1} \sum_{t=0}^{N_{cp}+N_{sc}-1} \left\|{\boldsymbol y}_q(t)-{\boldsymbol {\tilde x}}_q(t)\right\|_2^2.
\end{align}
Therefore, the readout weights are updated by the following closed-form expression,
\begin{align}
\label{output_weights}
{\boldsymbol W}_{tout} = \left(\left[{\boldsymbol S}_0^T,\cdots,{\boldsymbol S}_{Q-1}^T\right]^{T}\right)^{+}\left[{\boldsymbol {\tilde X}}_0^T,\cdots,{\boldsymbol {\tilde X}}_{Q-1}^T\right]^{T}
\end{align}
where $(\boldsymbol S)^+$ is the pseudo-inverse of $\boldsymbol S$, and ${\boldsymbol S}_q$ is stacked by the trajectory of the states as
\begin{align}
{\boldsymbol S}_q \triangleq [{\boldsymbol s}_q(0), {\boldsymbol s}_q(1), \cdots, {\boldsymbol s}_q(N_{sc}+N_{cp}-1)]^T.
\end{align}

In addition, due to the feedback nature of RC, there exists a lag-effect on the generated state response \cite{lukovsevivcius2009reservoir}. A delay parameter can be introduced in the learning process such that the following slightly revised training set is utilized\cite{holzmann2010echo},
\begin{align*}
d^{(p)}  \triangleq\big( [{\boldsymbol X}_0,\cdots ,{\boldsymbol X}_{Q-1}, {\boldsymbol O}_{N_r\times p}]^T, [ {\boldsymbol O}_{N_r\times p}, {\boldsymbol {\tilde X}}_0,\cdots ,{\boldsymbol {\tilde X}}_{Q-1},]^T\big)
\end{align*}
where $p \in [0,P]$ is the aforementioned delay parameter, and $\boldsymbol O$ represents the zero matrix. 
The time domain RC is trained with different values of $p$ with  $p^{\star}$ being the one generating the minimal objective value defined in (\ref{time_objective}). 
This input-output delay offset $p^{\star}$ will be used to configure the RC for testing.
Overall, the time domain RC-based MIMO-OFDM symbol detection can be summarized in Algorithm~\ref{algorithm0} where the parameter $P$ can be experimentally set as the maximum delay of the underlying channel. 

\begin{algorithm}[!ht]
	\caption{Time domain RC-based MIMO-OFDM symbol detection}
	\label{algorithm0}
	\begin{algorithmic}
		\REQUIRE $\{d_q\}_{q=0}^{Q-1}$
		\ENSURE ${\boldsymbol W}_{tout}$, $p^{\star}$
		\FOR{$p$ from $0$ to $P$}
		\STATE Generate $p$-delayed training set $d^{(p)}$
		\STATE Generate the state matrix $\{\boldsymbol S_{q}^{(p)}\}_{q=0}^{Q-1}$ according to the dynamics equation (\ref{Dynamic_Eq})
		\STATE Calculate the output weights ${\boldsymbol W}_{tout}^{(p)}$ using (\ref{output_weights}) and the objective value $Obj_p$ using (\ref{time_objective})
		\ENDFOR{}
		\STATE Find the optimal $p^{\star} =\arg\min_p Obj_p$ and ${\boldsymbol W}_{tout} = {\boldsymbol W}_{tout}^{p^{\star}}$
	\end{algorithmic}
\end{algorithm}

\subsection{Time-Frequency RC -- RC with Structural Information}
\begin{figure}[!h] 
	\centering
	\includegraphics[width=0.85\linewidth, height = 0.3\linewidth]{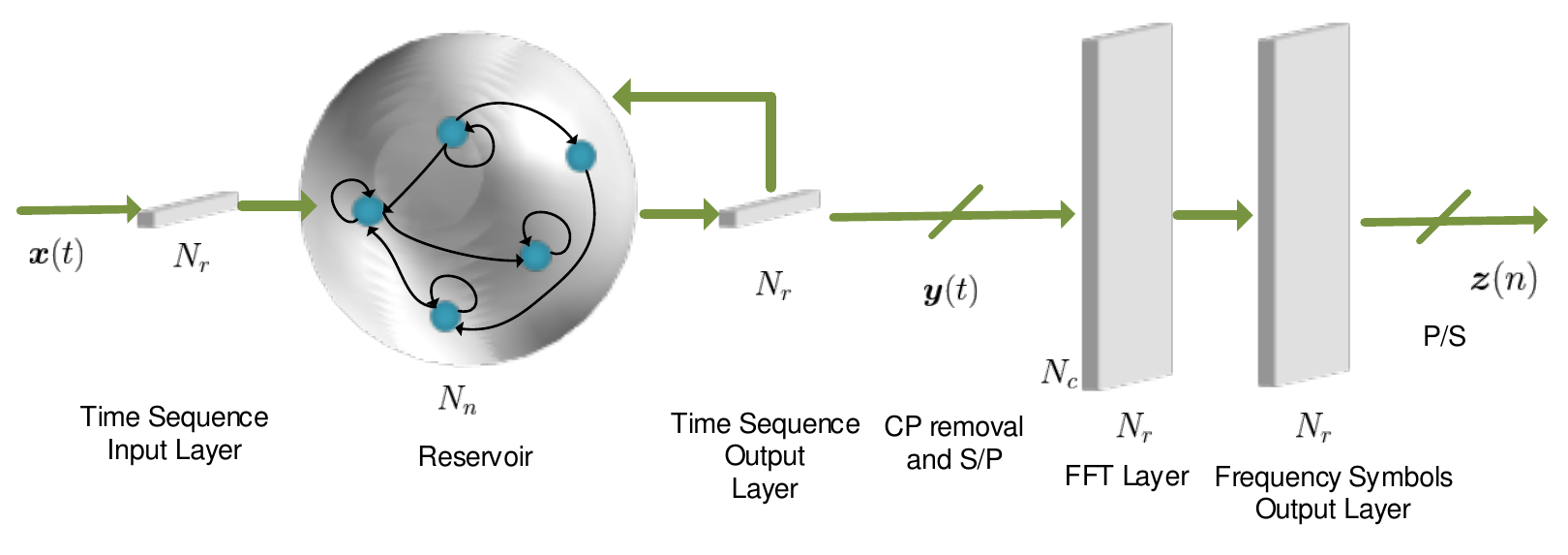}
	\caption{RCNet: Time-Frequency RC}
	\label{T_F_RC}
\end{figure}
Since the time domain RC focuses only on the time domain received signal without processing the frequency domain signal, it is clear that it does not take advantage of the structural information of the underlying OFDM signalization. 
In OFDM systems, the training signal is sent over the frequency domain and the frequency domain provides a much cleaner view of the transmitted signal due to the OFDM signalization.
Therefore, it is desirable for the NN-based symbol detector to conduct detection tasks on the frequency domain.
Meanwhile, the underlying NN-based symbol detector should also take advantage of the temporal correlation in the time domain.

To achieve these goals, we introduce the time-frequency RC whose structure is shown in Fig.~\ref{T_F_RC}. 
Following the same reasoning as the time domain RC, the reservoir first generates high dimensional response signals according to the input. 
The response signals are mapped to the desired signals through a time domain layer, a fixed FFT layer, and a frequency domain layer. Accordingly, the time domain layer output is the same as that in (\ref{time_output}). 
After removing the CP and conducting an FFT on $\{{\boldsymbol y}_q(t)\}_{t=N_{cp}}^{N_{sc}+N_{cp}-1}$, the obtained frequency domain symbols are denoted as $\{{\boldsymbol {\tilde y}}_n(n)\}_{n=0}^{N_{sc}-1}$. 
Subsequently, the frequency domain layer output is defined as
\begin{align}
{\boldsymbol {\boldsymbol {\tilde z}}}_q(n) = {\boldsymbol {\tilde y}}_n(n){\text {diag}}\left({\boldsymbol w}_{fout}(n)\right)
\end{align}
where ${\boldsymbol w}_{fout}(n)$ is the weight specified at the $n$th sub-carrier; $\text {diag}(\boldsymbol w)$ represents a diagonal matrix which has $\boldsymbol w$ as the main diagonal elements. 
Furthermore, we set the magnitude of each entry of ${\boldsymbol w}_{fout}(n)$ as $1$. 
This allows the introduced frequency domain layer to compensate for the residual phase error after the time domain processing. 
It is important to note that the frequency domain processing can effectively tune the delay parameter discussed in Section~\ref{subsec:TRC}.
This is because the FFT layer converts a shift in the time domain to a phase variation in the frequency domain. 
Accordingly, the added output layers of the time-frequency RC essentially leverage the structural information of the MIMO-OFDM signal.

The learning objective of the time-frequency output layer is
\begin{equation}
\label{time_frequency_RC_weights_learning}
\begin{aligned}
    \min_{\substack{{\boldsymbol W}_{tout}\\ \{{\boldsymbol w}_{fout}(n)\}_{n = 0}^{N_{sc}-1}}}& \sum_{q=0}^{Q-1} \sum_{n=0}^{N_{sc}-1} \left\|{\boldsymbol z}_q(n)-{\boldsymbol {\tilde z}}_q(n)\right\|_2^2 \\
    s.t. \quad &{\text{diag}}\left(\left|{\boldsymbol w}_{fout}(n)\right|\right) = {\boldsymbol I} \\
    \forall\quad& {n = 0,\cdots, N_{sc}-1}
\end{aligned}.
\end{equation}
which is equivalent to the time domain objective function defined in (\ref{time_objective}). In order to seek a proper solution to the above problem, we resort to alternative least squares (ALS) as the solver to obtain closed-form updating rules for ${\boldsymbol W}_{tout}$ and ${\boldsymbol w}_{fout}(n)$. The detailed derivation can be found in Appendix \ref{appendix_t_f_learning} with their closed-form updating rules outlined in (\ref{time_domain_W}) and (\ref{frequency_domain_W}) respectively. 
The learning algorithm of the time-frequency RC is summarized in Algorithm~\ref{algorithm1}.
\begin{algorithm}[!ht]
	\caption{Time-frequency RC based MIMO-OFDM symbol detection}
	\label{algorithm1}
	\begin{algorithmic}[Forward and Back Propagation on Weights] 
		\REQUIRE $\{d_q\}_{q=0}^{Q-1}$
		\ENSURE ${\boldsymbol W}_{tout}$, $\{{\boldsymbol w}_{fout}(n)\}_{n = 0}^{N_{sc}-1}$
		\STATE Generate the state matrix $\{\boldsymbol S_{q}^{(p)}\}_{q=0}^{Q-1}$ according to the dynamics equation (\ref{Dynamic_Eq})
		\STATE Initialize ${\boldsymbol w}_{fout}(n) = {\boldsymbol 1}$, $\forall n = 1,\cdots, N_{sc}$
		\WHILE{(\ref{time_frequency_RC_weights_learning}) does not converge }
		\STATE Update ${\boldsymbol W}_{tout}$ using (\ref{time_domain_W})
		\STATE Update ${\boldsymbol w}_{fout}(n)$ using (\ref{frequency_domain_W})
		\ENDWHILE{}
	\end{algorithmic}
\end{algorithm}

Note that there are certain implementation-related issues that need to be clarified when fully connected layers are employed in the frequency domain. 
In this case, the learning rule of the output layers becomes
\begin{align*}
\label{full_layers_frequency}
\min_{\substack{
 {\boldsymbol W}_{tout}\\
\{{\boldsymbol W}_{fout}(n)\}_{n=0}^{N_{sc}-1}}
}\sum_{n=0}^{N_{sc}-1} \left\|{\boldsymbol {\tilde F}}(n){\boldsymbol S}{\boldsymbol W}_{tout}{\boldsymbol W}_{fout}(n)-{\boldsymbol { Z}(n)}\right\|_F^2 
\end{align*}
Given ${\boldsymbol W}_{tout}$, the updating rule for ${\boldsymbol W}_{fout}(n)$ is given by
\begin{align}
    {\boldsymbol W}_{fout}(n) = ({{\boldsymbol {\tilde F}}(n){\boldsymbol S}{\boldsymbol W}_{tout}})^+{\boldsymbol Z}(n).
\end{align}
Given a ${\boldsymbol W}_{fout}(n)$, ${\boldsymbol W}_{tout}$ can be updated by solving a Sylvester's equation which can introduce heavy computational load. 
Meanwhile, this extension brings more parameters to learn which can lead to overfitting since the size of the training set is usually limited in practical systems. 


\section{RCNet -- Stacking RCs for a Deep Network}
\label{Layered_RC0}
\begin{figure*}[h!]
	\centering
	\includegraphics[width=1\linewidth, height = 0.2\linewidth]{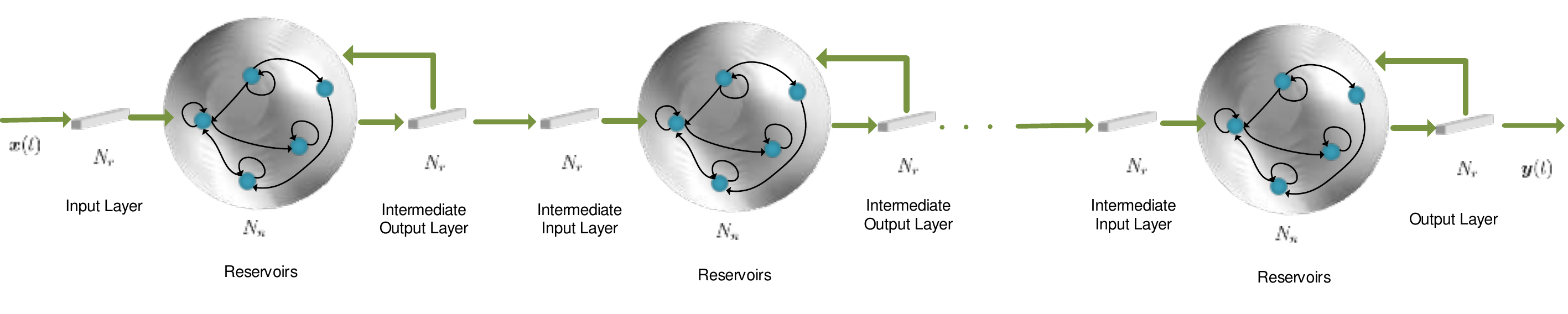}
	\caption{RCNet: Deep Time RC}
	\label{Layered_RC}
\end{figure*}
We now introduce RCNet by stacking multiple RCs into a ``deep'' RC network. 
Intuitively, this deep structure can be interpreted as decomposing different levels of interference cancellation for the received MIMO-OFDM signal. 
Based on the discussion in Section~\ref{Time_Frequency_RC}, the basic building block of RCNet can be either the time domain RC and the time-frequency RC.

When the building block is the time domain RC, RCNet can be constructed by stacking them into a deep structure (Deep Time RC) as shown in Fig.~\ref{Layered_RC}. 
Let $L$ denote the total number of building block in this RCNet. Given the training set $d^{(p)}$, for the $l$th RC, the state equations are
\begin{align*}
{\boldsymbol s}_q^{(l)}(t+1) = f\left({\boldsymbol s}_q^{(l)}(t){\boldsymbol W}_s^{(l)}+{\boldsymbol y}_q^{(l-1)}(t){\boldsymbol W}_{in}^{(l)}+{\boldsymbol {\tilde x}_q}(t){\boldsymbol W}_{fb}^{(l)}+{\boldsymbol n}(t)\right)
\end{align*}
where the superscript $(l)$ represents the $l$th RC; ${\boldsymbol y}^{(l)}$ follows the output equation from the previous layer which is defined in (\ref{time_output}). 
The output layer and intermediate output layers of this RCNet are learned sequentially, i.e. the intermediate output layer closest to the input is learned first; the next RC is learned based on the results generated by the previous one. 
The input of the $l$th RC is the output of the $(l-1)$th RC after training. 
The teacher forcing for different RCs is the same and the final output of this RCNet generates an estimate of the desired MIMO-OFDM symbol, i.e. ${\boldsymbol y}_q^{(L)}(t)$. 
The learning algorithm of the deep time RC is summarized in Algorithm~\ref{algorithm3}. 
Such a learning method can be interpreted through the lens of the boosting framework~\cite{freund1995boosting}, which states that by sharing learned features among a set of weak learners, their ensemble can result in a stronger learning ability.
\begin{algorithm}[!h]
	\caption{RCNet: Deep Time RC based MIMO-OFDM Symbol Detection}
	\label{algorithm3}
	\begin{algorithmic}
		\REQUIRE $\{d_q\}_{q=0}^{Q-1}$
		\ENSURE $\{{\boldsymbol W}_{tout}^{(l)}\}_{l=0}^{L-1}$, $\{p^{(l)}\}_{l = 0}^{L-1}$
		\STATE $\{{\boldsymbol x}_q^{(o)}(t)\}_{t=0}^{N_{cp}+N_{sc}-1}=\{{\boldsymbol x}_q(t)\}_{t=0}^{N_{cp}+N_{sc}-1}$
		\FOR{$l$ from $0$ to $L-1$}
		\STATE $d_q(l) \triangleq \Big(\{{\boldsymbol x}_q^{(l)}(t)\}_{t=0}^{N_{cp}+N_{sc}-1},\{{\boldsymbol {\tilde x}}_q(t)\}_{t=0 }^{N_{cp}+N_{sc}-1}\Big) $
		\FOR{$p$ from $0$ to $P$}
		\STATE Generate $p$-delayed training set $d^{(p)}(l)$
		\STATE Generate the state matrix $\{\boldsymbol S_{q}^{(p)}\}_{q=0}^{Q-1}$ according to the state equation (\ref{Dynamic_Eq})
		\STATE Calculate the output weights ${\boldsymbol W}_{tout}^{(p)}$ using (\ref{output_weights}) and the objective value $Obj_p$ using (\ref{time_objective})
		\ENDFOR{}
		\STATE Find the optimal $p^{(l)} =\arg\min_p Obj_p$ and ${\boldsymbol W}_{tout}^{(l)} = {\boldsymbol W}_{tout}^{p^{(l)}}$
        \STATE Use the learned ${\boldsymbol W}_{tout}^{(l)}$ and $p^{(l)}$ to generate $\{{\boldsymbol x}_q^{(l+1)}(t)\}_{t=0}^{N_{cp}+N_{sc}-1}$
		\ENDFOR{}
	\end{algorithmic}
\end{algorithm}

Note that the number of RC components in the RCNet is another tunable parameter. 
A validation set can be utilized for determining the proper value of $L$. 
We can simultaneously test the validation error while increasing $L$: Once the validation error stops decreasing, we can stop increasing $L$. 
Furthermore, the number of neurons in each RC can also be configured differently. 
How to optimize the number of neurons in each RC to achieve the best generalization performance can be treated as a future work.

\begin{figure*}[h]
	\centering
	\includegraphics[width=1\linewidth, height = 0.175\linewidth]{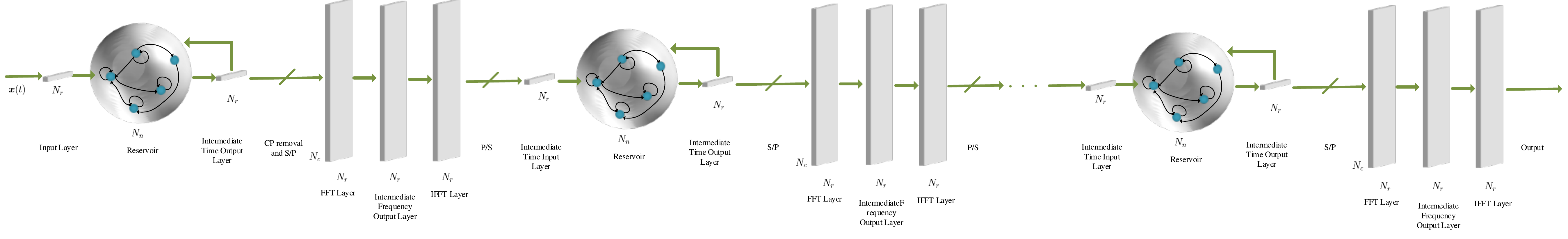}
	\caption{RCNet: Deep Time-Frequency RC structure}
	\label{Layered_RC2}
\end{figure*}

Similarly, we can change the building block of RCNet from time domain RC to the newly introduced time-frequency RC. 
The corresponding structure of the RCNet (Deep Time-Frequency RC) can be shown in Fig. \ref{Layered_RC2}. 
It is expected the Deep Time-Frequency RC will provide better performance than the Deep Time RC since it takes advantage of the structural information of OFDM signalization.
Furthermore, it is important to note that compared to the Deep Time RC version of the RCNet, additional IFFT layers need to be added between two adjacent time-frequency RCs to transform the former frequency domain output to a time domain signal for subsequent processing. 
A similar learning algorithm for this RCNet (Deep Time-Frequency RC) is summarized in Algorithm~\ref{algorithm4}.

\begin{algorithm}[!ht]
	\caption{RCNet: Deep Time-Frequency RC based MIMO-OFDM Symbol Detection}
	\label{algorithm4}
	\begin{algorithmic}[Forward and Back Propagation on Weights] 
		\REQUIRE $\{d_q\}_{q=0}^{Q-1}$
		\ENSURE $\{{\boldsymbol W}_{tout}^{(l)}\}_{l=0}^{L-1}$, $\{p^{(l)}\}_{l = 0}^{L-1}$
		\STATE $\{{\boldsymbol x}_q^{(o)}(t)\}_{t=0}^{N_{cp}+N_{sc}-1}=\{{\boldsymbol x}_q(t)\}_{t=0}^{N_{cp}+N_{sc}-1}$
		\FOR{$l$ from $0$ to $L-1$}
		\STATE $d_q(l) \triangleq \Big(\{{\boldsymbol x}_q^{(l)}(t)\}_{t=0}^{N_{cp}+N_{sc}-1},\{{\boldsymbol {\tilde x}}_q(t)\}_{t=0 }^{N_{cp}+N_{sc}-1}\Big) \cong \Big(\{{\boldsymbol x}_q^{(l)}(t)\}_{t=0}^{N_{cp}+N_{sc}-1},\{{\boldsymbol z}_q(n)\}_{n=0 }^{N_{sc}-1}\Big)$
		\STATE Generate the state matrix $\{\boldsymbol S_{q}^{(p)}\}_{q=0}^{Q-1}$ according to the state equation (\ref{Dynamic_Eq})
		\STATE Initialize ${\boldsymbol w}_{fout}(n) = {\boldsymbol 1}$, $\forall n = 1,\cdots, N_{sc}$
		\WHILE{(\ref{time_frequency_RC_weights_learning}) does not converge }
		\STATE Update ${\boldsymbol W}_{tout}$ using (\ref{time_domain_W})
		\STATE Update ${\boldsymbol w}_{fout}(n)$ using (\ref{frequency_domain_W})
		\ENDWHILE{}
		\STATE Use the learned ${\boldsymbol W}_{tout}$ and ${\boldsymbol w}_{fout}(n)$ to generate $\{{\boldsymbol x}_q^{(l+1)}(t)\}_{t=0}^{N_{cp}+N_{sc}-1}$
		\ENDFOR{}
	\end{algorithmic}
\end{algorithm}

\section{Performance Evaluation}
\label{Numerical_Experiments}
In this section, we provide performance evaluations for the introduced RCNet framework under relevant scenarios. 
The modulation scheme used to generate $z(n)$ is set to be $16$-QAM. The parameters in performance evaluation are configured as the follows: $N_r = 4$, $N_t = 4$, $N_{sc} = 1024$, $N_{cp} = 160$, $Q = 4$, and $N_d = 13$. 
Note that in this case the training overhead is only $23.5\%$ which is completely different from most other NN-based detection methods that use a prohibitively high training overhead.
The channel model adopted in the evaluation is the Winner II channel model defined in~\cite{meinila2009winner}, where the transmitter and receiver are configured with uniform linear arrays having half-wavelength antenna spacing.
Furthermore, the communication scenario is chosen to be the outdoor-to-indoor case. 

Recall from Fig.~\ref{OFDM_resource_grid} that the first $Q$ OFDM symbols are the training set and next $N_d$ OFDM symbols are the data symbols. 
Therefore, the first $Q$ OFDM symbols will be used to train the RCNet while the bit error rate (BER) will be evaluated using the testing set (data symbols).
This training and testing procedure will be conducted for $100$ consecutive sub-frames. 
The number of neurons for each layer of the RCNet (the component of the shallow RC) is set as $128$. 
Furthermore, a time window is added to the input layer of each RC unit as suggested in~\cite{zhou2019}, where the length is set as $128$. 
The state transition matrix ${\boldsymbol W}_s$ is generated randomly to satisfy the echo state property~\cite{jaeger2001echo}, where the spectral radius is chosen to be smaller than $1$. 
The input weight ${\boldsymbol W}_{in}$ is generated randomly from a uniform distribution. 
In our evaluation, adding teacher forcing does not show a difference in performance. 
Therefore, ${\boldsymbol W}_{fb}$ is set to zero. 

\subsection{BER Performance under Tx Non-linearity}
\label{ber_tx_nonlinearity}
\begin{figure}
\centering
\includegraphics[width=0.835\linewidth, height =0.625\linewidth]{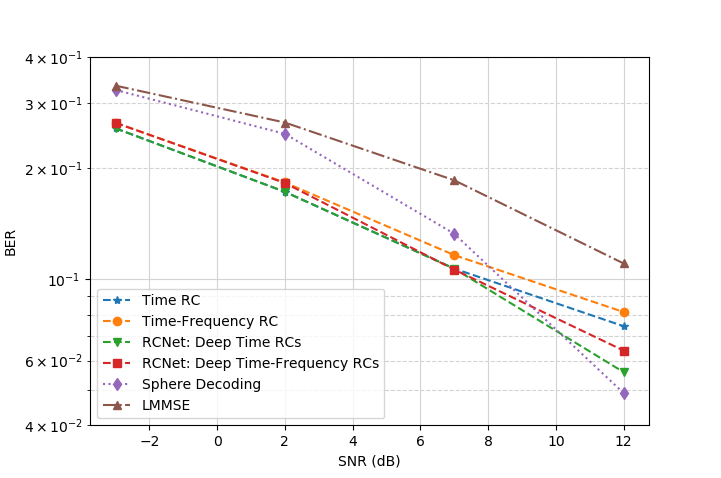}
\caption{Average BER for RCNet-based methods and conventional methods in PA linear region}
\label{ber1}
\end{figure}

\begin{figure}
\centering
\includegraphics[width=0.75\linewidth, height =0.625\linewidth]{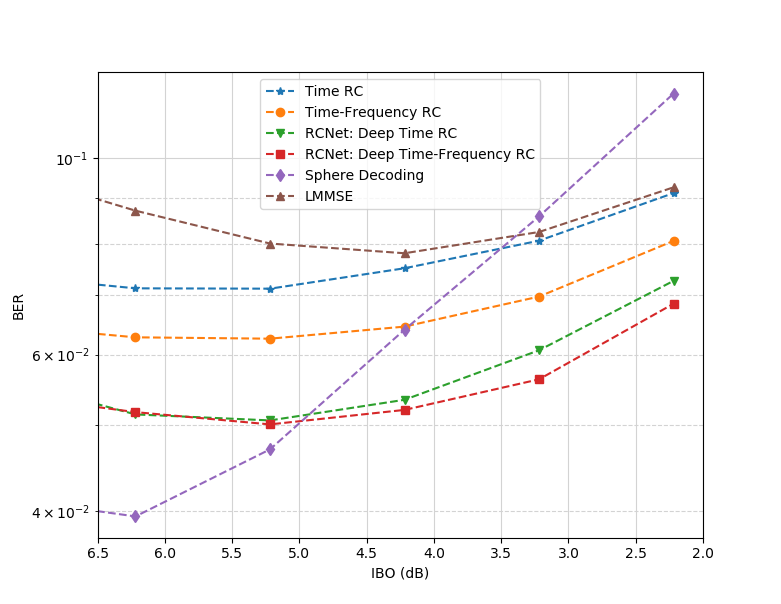}
\caption{Average BER for RCNet-based methods and conventional methods by varying IBO to operate in PA's non-linear region}
\label{ber2}
\end{figure}
To evaluate the symbol detection performance, we compare RCNets to shallow RC-based strategies as well as conventional methods. 
To incorporate the Tx non-linearity in our evaluation, the following RAPP model was adopted for PA,
\begin{align}
f(x) = {\frac{x}{\left[1+\left(\frac{|x|}{x_{sat}}\right)^{2\rho}\right]^{1/{2\rho}}}}
\end{align}
where $x$ represents the PA input signal, $\rho$ is the smoothing parameter, and $x_{sat}$ is the saturation level. 
When $x \ll x_{sat}$, we have $f(x)\approx x$ which means the PA output signal has no distortion compared with the input. 
In our evaluation, we set $\rho = 3$ and $x_{sat} = 1$. 
As a benchmark, two conventional symbol detection methods, namely linear minimum mean squared error (LMMSE) and sphere decoding (SD) \cite{ghasemmehdi2011faster}, are selected. 
Since these two methods rely on the knowledge of channel state information (CSI), we utilize LMMSE as the channel estimation strategy based on the $Q$ OFDM symbols of reference signals assuming the perfect knowledge of the noise variance and the linearity of the underlying wireless link.

We first consider the BER performance when the PA input power is in the linear region. 
In this case, the PA input power is backed off to be far from the PA's saturation region. 
Accordingly, the input back-off (IBO), which is defined as the ratio between the PA's saturation power to the input power, is chosen to be greater than $8$ ${\text {dB}}$. 
In Fig.~\ref{ber1} we show the bit error rate (BER) plotted as a function of the received signal to noise ratio (SNR) in dB for various symbol detection schemes.
From the results we can observe that all RC-based methods including the shallow structures and the RCNet can perform better than conventional methods when the transmission power is low.
This is because the estimated CSI is inaccurate in the low SNR regime resulting in poor performance of the conventional model-based methods. 
On the other hand, RC-based methods are able to learn the underlying features of the channel without explicitly relying on the underlying CSI.
Furthermore, it can seen from the results that the two versions of RCNet provide slightly performance improvement over its shallow counterparts demonstrating the benefits of the deep network structure. 

In Fig.~\ref{ber2}, we show the BER performance when the PA input power is close to the saturation region. 
In this case, the PA's output is distorted due to the compression effects. 
The distortion occurs when the peak-to-average-power-ratio (PAPR) of the PA input signal is higher than the value of the IBO, where the PAPR of an OFDM signal $x(t)$ is defined as $\|x(t)\|_{\infty}^2/\|x(t)\|_{2}^2$. 
In our evaluation, the signal's PAPR is controlled in the range from $6$ ${\text{dB}}$ to $9$ ${\text{dB}}$. 
Therefore, in order to investigate the BER performance under PA's compression effects, we evaluate the BER performance by choosing the IBO below $6.5$ $\text{dB}$ as shown in the figure. 
From the results we can clearly see that all RC-based methods perform relatively well when the IBO is low, especially when it is lower than $5$ dB. 
Note that the PA efficiency is substantially higher when it is operating at a low IBO. 
This clearly suggests that RC-based methods can provide an improvement in PA efficiency by compensating for the transmitted waveform distortion at the receiver.
Furthermore, the results in Fig.~\ref{ber2} demonstrate the benefits of structural information in the underlying NN design: the Time-Frequency RC performs significantly better than the Time RC and the RCNet with Deep Time-Frequency RC performs better than the RCNet with Deep Time RC.
This is because the newly introduced Time-Frequency RC does take advantage of the OFDM signalization in the design of the underlying network structure of RC to address \emph{Challenge 2}.
On the other hand, the evaluation results also clearly show the powers of the deep nature of the introduced RCNet to address \emph{Challenge 3} without additional training overhead: RCNet performs substantially better than its shallow counterparts with the same training overhead.

\subsection{BER Performance under Rx Non-linearity}
The non-linearity in the receiver stems primarily from the quantization of the received signal due to finite resolution analog-to-digital conversion. 
For a MIMO-OFDM signal, the in-phase and quadrature components are quantized by a pair of analog-to-digital converters (ADCs), where the input-output relation of the ADCs can be defined as
\begin{equation}
q(x) = \left\{
\begin{array}{lr}
\Delta\lceil x/\Delta \rceil-{\Delta/2}, & {\text{if}}|x| < A_{max} \\
A_{max}\cdot {\text{sign}}(x), & {\text{otherwise}}
\end{array}
\right.
\end{equation}
in which $x$ is the ADC's input, $\lceil \cdot \rceil$ is the ceiling function, $\Delta>0$ represents the quantization interval, and $A_{max}$ is the maximal amplitude of ADC such that $A_{max}=(2^n-1)\Delta/2$, where $n$ is the number of quantization bits. 
To be specific, when $\Delta = 2 A_{max}$, $q(x)$ represents a one-bit ADC. 
In the state-of-the-art, low-resolution ADCs are often utilized to digitize a signal with large bandwidth at high frequencies (e.g., mmWave bands and Terahertz bands) or to reduce the power consumption of the underlying ADCs. 

\begin{figure}
\centering
\includegraphics[width=0.9\linewidth, height =0.625\linewidth]{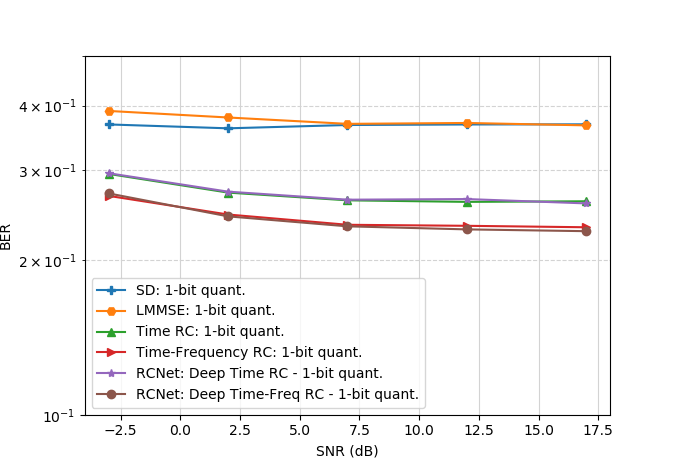}
\caption{ Average BER curves for RCNet-based methods and conventional methods using 1-bit resolution ADCs}
\label{quantization_results}
\end{figure}

\begin{figure}
    \centering
    \includegraphics[width=0.9\linewidth, height =0.625\linewidth]{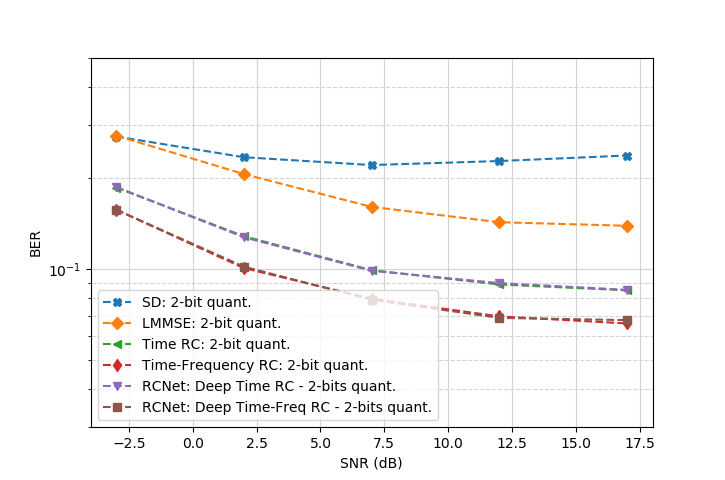}
    \caption{Average BER curves for RCNet-based methods and conventional methods using 2-bit resolution ADCs}
    \label{quantization_results_2bit}
\end{figure}

In Fig.~\ref{quantization_results} and Fig.~\ref{quantization_results_2bit}, we investigate the BER performance of uncoded MIMO-OFDM signals under $1$- and $2$-bit quantization respectively ($n=1$ and $n=2$) using the QPSK modulation as a function of the received SNR. 
Since only $1$-bit and $2$-bit ADCs are used in these cases the resulting quantization errors are usually large.
From the figures we can see that conventional methods are very sensitive to quantization errors where the SD strategy completely collapses in both cases.
On the other hand, all RC-based methods can handle the large quantization errors very well showing the benefits of adopting learning-based approach for MIMO-OFDM symbol detection. 
It is important to note that the RC-based strategies outperform the conventional LMMSE and SD in all SNRs of interests.
Furthermore, a substantial gain can be achieved from Time-Frequency RC to Time RC showing the importance of incorporating structural information in the underlying NN design for MIMO-OFDM symbol detection.
As for the comparison between the shallow RCs and the RCNets, we see very marginal BER performance improvement. 
This might be due to the fact that the shallow RC already takes the best advantage of the available limited information for symbol detection where additional iteration does not provide performance improvement.
Identifying the detailed reasons behind this phenomenon can be treated as a future extension of this work.


\subsection{Learning Convergence of RCNet}
\begin{figure}
\centering
\includegraphics[width=0.8\linewidth, height =0.625\linewidth]{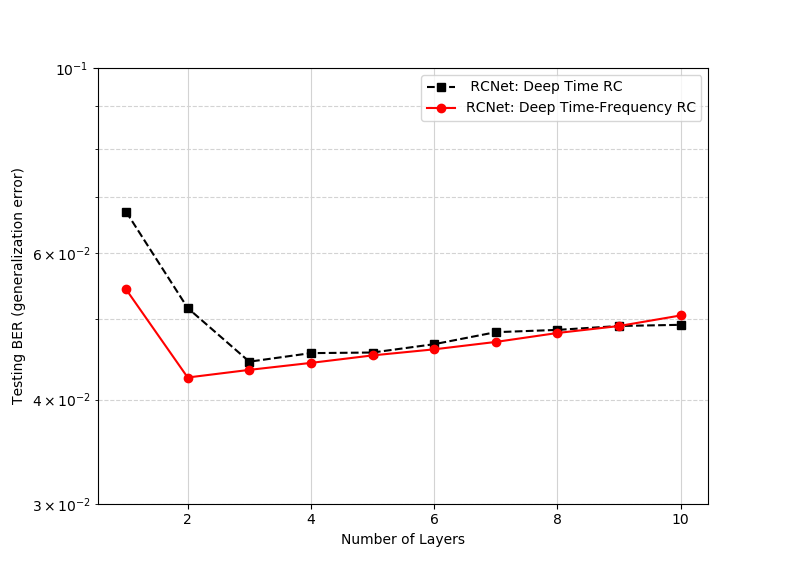}
\caption{Testing BER (Generalization Error) for RCNet as a function of $L$}
\label{generalization}
\end{figure}

The evaluation results presented in previous sections show that RC-based methods are effective in the low SNR regime and under the effects of transmission and reception non-linearities. 
In these evaluations, we set $L = 3$ for RCNet. 
Intuitively, we can increase $L$ to yield better training performance, however, a higher $L$ may cause over-fitting. 
Fig.~\ref{generalization} shows the testing BER of RCNet as a function of $L$ considering only the transmit non-linearity.
As shown in the figure, increasing $L$ does ensure a decrease in the generalization error.

We now evaluate the convergence behavior of RCNet under transmit non-linearity . 
The $E_b/N_0$ is set to be $15$ dB and we train different RCs under the same channel realization. 
The objective function used for tracing the number of iterations is defined in~(\ref{time_objective}) for the Time RC and in (\ref{time_frequency_RC_weights_learning}) for the Time-Frequency RC.
\begin{figure}
\centering
\includegraphics[width=0.8\linewidth, height =0.625\linewidth]{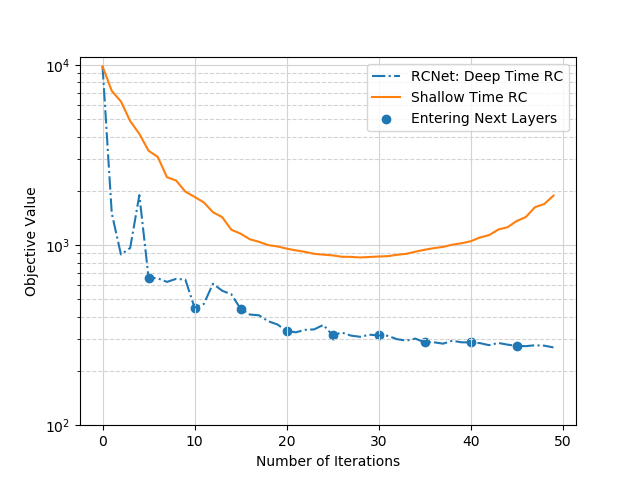}
\caption{Learning curves of the Shallow Time RC and RCNet (Deep Time RC)}
\label{learning_curve_cascade1}
\end{figure}
Fig.~\ref{learning_curve_cascade1} shows the learning curve of the time domain RCs where each iteration corresponds a fixed delay parameter. 
For the delayed training of RCNet, we set $P = 5$ and choose the delay parameters uniformly from $0$ to $N_{cp}$ in Algorithm~\ref{algorithm3}. 
Therefore, the training for each RC layer of the RCNet requires $5$ iterations and a total of $L = 10$ layers is considered in the figure.
For the shallow time RC, we set $P = 50$ in Algorithm~\ref{algorithm1} meaning that the resolution used in the search for the optimum delay is finer for the case of shallow RC than that for the case of RCNet.
Therefore, there are a total of $50$ iterations for the shallow Time RC.
From Fig.~\ref{learning_curve_cascade1} we clearly observe that a finer delay parameter cannot lead to a lower objective value, while adding extra layers of RCs on top of the shallow RC can decrease the objective value.
This clearly demonstrates the benefits of the deep structure.

\begin{figure}
\centering
\includegraphics[width=0.8\linewidth, height =0.625\linewidth]{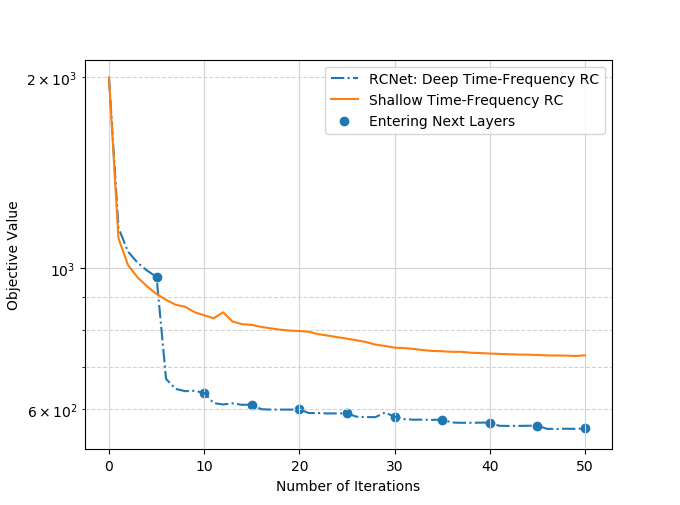}
\caption{Learning curves of the Shallow Time-Frequency RC and RCNet (Deep Time-Frequency RC)}
\label{learning_curve_cascade2}
\end{figure}
A similar conclusion can be drawn by observing the learning curve of the Time-Frequency RC and RCNet as depicted in Fig.~\ref{learning_curve_cascade2}.
For RCNet (Deep Time-Frequency RC), we fix the number of iterations of the ALS for solving each RC layer to be $5$ in Algorithm~\ref{algorithm4}.
Therefore, as in Fig.~\ref{learning_curve_cascade1}, the training for each RC layer of the RCNet requires $5$ iterations and a total of $L = 10$ layers is considered in the figure.
As shown in Fig.~\ref{learning_curve_cascade2}, the objective value decreases significantly by adding one extra RC layer. 
Compared with the shallow Time-Frequency RC, the fitting error of RCNet (Deep Time-Frequency RC) can be significantly smaller.

\begin{figure}
\centering
\includegraphics[width=0.8\linewidth, height =0.625\linewidth]{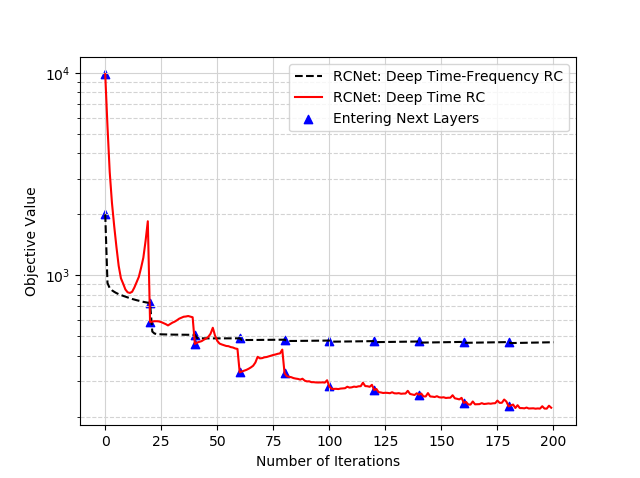}
\caption{Learning curve of the Deep Time RC and Deep Time-Frequency RC structures in RCNet}
\label{learning_curve_cascade3}
\end{figure}

The comparison between the two RCNets (Deep Time RC and Deep Time-Frequency RC) is presented in Fig.~\ref{learning_curve_cascade3}.
In this comparison, we set the number of iterations of each RC layer to be $20$ with $L$ changes from $1$ to $10$.
From the figure we can see that the RCNet with Deep Time-Frequency RC yields a stable objective value where increasing $L$ no longer helps in deceasing the training error.
On the other hand, increasing $L$ always helps to reduce the training error for the RCNet with Deep Time RC.
Considering the generalization error on the testing set shown in Fig.~\ref{generalization}, it is clear that the stability characteristics of the learning curve for RCNet with Deep Time-Frequency RC provides an efficient way to determine a good value of $L$ to avoid overfitting issues.
This is another important advantage of the Deep Time-Frequency RC.

\subsection{Comparison with Other NNs}
\label{subsec:Comparison}
\begin{table}
\centering
\caption{Bit Error Rate (BER) Comparison with alternate NN-based methods (PA IBO = $2.2$ $dB$)}
\begin{tabular}{lccccc} 
\textbf{Detection Framework}           & \textbf{Training Symbols} & \textbf{$\eta$ (\%)}     & \textbf{Training Epochs} & \textbf{Training BER} & \textbf{Testing BER}  \\ 
\hline
\textbf{Time RC~\cite{zhou2019}}                       & \multirow{4}{*}{$4$}        & \multirow{4}{*}{$23.5$} & \multirow{4}{*}{N/A}     & \num{2e-3}                 & \num{9e-2}                  \\
\textbf{Time-Frequency RC}             &                           &                       &                          & \num{8e-3}                 & \num{8e-2}                  \\
\textbf{RCNet: Deep Time RC}           &                           &                       &                          & \num{1.4e-3}                & \num{7.3e-2}                 \\
\textbf{RCNet: Deep Time-Frequency RC} &                           &                       &                          &  \num{3.5e-2}                & \num{6.9e-2}                 \\ 
\hline
\textbf{LSTM}                          & \multirow{3}{*}{$4$}        & \multirow{3}{*}{$23.5$} & \multirow{3}{*}{$300$}     & \num{3e-2}                  & \num{4.7e-1}                  \\
\textbf{Bi-LSTM}                       &                           &                       &                          & \num{4.6e-5}             & \num{4.7e-1}                  \\
\textbf{GRU}                           &                           &                       &                          & \num{6e-2}                 & \num{4.7e-1}                  \\ 
\hline
\textbf{LSTM}                          & \multirow{3}{*}{$10$}       & \multirow{3}{*}{$43.5$} & \multirow{3}{*}{$300$}     & \num{2.2e-1}                      & \num{4.7e-1}                  \\
\textbf{Bi-LSTM}                       &                           &                       &                         & \num{5.1e-2}                     & \num{4.7e-1}                  \\
\textbf{GRU}                           &                           &                       &                          & \num{2.7e-1}                      & \num{4.8e-1}                  \\ 
\hline
\textbf{LSTM}                          & \multirow{3}{*}{$20$}       & \multirow{3}{*}{$60.6$} & \multirow{3}{*}{$300$}     & \num{2e-1}                      & \num{4e-1}                  \\
\textbf{Bi-LSTM}                       &                           &                       &                          & \num{9.9e-2}                      & \num{4e-1}                  \\
\textbf{GRU}                           &                           &                       &                          & \num{2.5e-1}                      & \num{4.1e-1}                  \\ 
\hline
\textbf{LSTM}                          & \multirow{3}{*}{$500$}      & \multirow{3}{*}{$97.5$} & \multirow{3}{*}{$30$}      & \num{4.5e-1}                  & \num{4.4e-1}                  \\
\textbf{Bi-LSTM}                       &                           &                       &                          & \num{4.4e-1}                  & \num{4.3e-1}                  \\
\textbf{GRU}                           &                           &                       &                          & \num{4e-1}                   & \num{4.4e-1}                  \\ 
\hline
\textbf{LSTM}                          & $5000$                     & $99.7$                  & $30$                       &  \num{4.4e-1}                     & \num{4.3e-1}                      \\ 
\hline
\textbf{LMMSE}                         & \multirow{2}{*}{$4$}                          & \multirow{2}{*}{$23.5$}                      & \multirow{2}{*}{N/A}                         & \multirow{2}{*}{N/A}                      & \num{9.2e-2}                 \\
\textbf{Sphere Decoder}                &                           &                       &                          &                       & \num{1.3e-1}                  \\ 
\hline
\textbf{DetNet~\cite{samuel2019learning}:}                        &                           &                         &                           &                        &                       \\
\textbf{Training with Perfect CSI}                       & \multirow{2}{*}{$3000$}                          & \multirow{2}{*}{$83.5$}                    & \multirow{2}{*}{\num{2e5}}                             & \num{5.7e-5}                                         & \num{1.2e-1}                                  \\
\textbf{Training with Estimated CSI}     &       &       &       & \num{8.8e-3}      & \num{7e-2} \\
\end{tabular}
\label{RNN_comparison}
\end{table}
In Table~\ref{RNN_comparison}, we present the performance comparison between the RC-based symbol detectors (shallow RCs and RCNets) and symbol detectors constructed using alternate NN architectures. 
To be specific, other popular RNN architectures are considered for performance comparison against RCNet.
The variants include long short-term memory (LSTM), bidirectional-LSTM (Bi-LSTM) and gated recurrent unit (GRU) that are more robust against the short-term memory issue in vanilla RNNs. 
Bi-LSTM is an extension of traditional LSTMs, training on both the original input sequence as well as its reversed copy leading to doubling of the training time for a given size of the reference signal set. 
To keep the comparison fair, the number of units in the recurrent hidden layer for the LSTM, Bi-LSTM and GRU structures is set to $128$, in line with the $128$ neurons used in each layer of the RCNet. 
In this comparative analysis, only the PA non-linearities in the transmitter are exercised by choosing a low input back-off (IBO) of $2.2$ $\text{dB}$ while ignoring quantization effects of the ADCs in the receiver.
It can be seen that in the case of all three RNN-variants, even a prohibitively large training set of $500$ reference OFDM symbols, amounting to a reference signal (training) overhead of $\eta = 97.5\%$, is not sufficient to achieve an acceptable BER for data transmission. 
On the other hand, with only $4$ reference OFDM symbols, i.e., a reference signal overhead of $\eta = 23.5\%$, all four RCNet detection methods achieve better BER performance than conventional detection methods (SD and LMMSE) listed in Table~\ref{RNN_comparison}. 
Note that SD performs worse than LMMSE since we are operating in the non-linear region of the PA, as can be confirmed from the results in Fig.~\ref{ber2}.

As a benchmark to compare RCNet against, we also evaluate another deep learning-based symbol detection scheme~\textit{DetNet} presented in \cite{samuel2019learning}. 
A key requirement of DetNet is the availability of perfect CSI during testing for MIMO symbol detection. 
However, perfect CSI is either infeasible or extremely costly to obtain in practical wireless systems. 
In order to provide a fair comparison of DetNet against RCNet which does not rely on CSI for detection, we evaluate DetNet's performance in the following two cases: 1) Training with perfect CSI, and 2) Training with estimated CSI. 
In Case 1) we assume perfect CSI is available in the training phase of DetNet. 
Note that the perfect training CSI assumption is valid for DetNet since the amount of training data is abundant for the receiver to obtain the close-to-perfect CSI.
On the other hand, during the testing phase of DetNet, like in the case of LMMSE and SD, we assume the LMMSE channel estimator is adopted to obtain the underlying CSI estimate. 
Accordingly, the estimated CSI is utilized in the testing phase.
In Case 2) we assume the LMMSE estimator is also adopted in the training phase of DetNet to make sure the training and testing environments are the same.
It is important to note that in both cases the training/channel estimation overhead is the same: the training overhead in the training phase + CSI estimation overhead in the testing phase. 
Further realizing that a PA backoff of $2.2$ $\text{dB}$ maps to a received SNR of $17$ $\text{dB}$ for RCNet, the same SNR value is used to train and test DetNet with its default configurations of parameters such as number of training iterations, learning rate and batch sizes.
Table~\ref{RNN_comparison} clearly suggests that RCNet can outperform DetNet in both cases.
To be specific, training with perfect CSI, DetNet can reach a testing BER of $0.12$ with a training overhead of $83.5\%$ (a training overhead of $60\%$ and a pilot overhead of $23.5\%$ for CSI estimation).
On the other hand, the testing BER for DetNet is $0.07$ when trained with estimated CSI.
Both results are higher than $0.069$, the testing BER of RCNet (Deep Time-Frequency RC) with a training overhead of $23.5\%$. 

Overall, from the comparison with other RNN-based symbol detection strategies we can clearly see that RCNet offers advantages of requiring very limited training overhead to provide good performance of MIMO-OFDM symbol detection tasks. 
Compared with conventional model-based symbol detection strategies, we can see RCNet provides advantages such as being robust to model mismatch and RF non-linearities (transmitter and/or receiver).
On the other hand, the state-of-art NN-based symbol detection strategy, \textit{DetNet}, is also robust to model mismatch and RF non-linearities as observed from the comparison between DetNet and conventional model-based strategies.
However, it requires extensive amount of training data, time, and computational resources compared to the RCNet.
\section{Conclusion and Future Work}
\label{Conclusion}
RC-based symbol detectors have been introduced for MIMO-OFDM systems to work under very limited training sets.
This paper introduced a deep RNN-based network called \emph{RCNet} to 1) incorporate structural information of the OFDM signalization, and 2) deepen the original shallow RC-based symbol detection strategies to further improve the detection performance of RC-based symbol detectors.
Incorporating structural information has been achieved through the invention of Time-Frequency RC where the learning is done both at the time and frequency domains to take advantage of the time-frequency structure of the underlying OFDM signals.
Meanwhile, the deep nature of RCNet has been achieved by extending a shallow RC structure to a deep RC structure in the following two ways: cascading time domain RCs and cascading Time-Frequency RCs.
The associated learning algorithms have been developed for each of the extensions and extensive evaluation has been conducted. 
Experimental results showed that RCNet can outperform conventional methods using the same limited training set under the non-linear RF effects of the wireless link demonstrating the effectiveness of incorporating structural information as well as deepening RCs for symbol detection tasks.  

An important area of exploration for future research is how to determine the optimal $L$, that is, the number of RCs in RCNet for MIMO-OFDM symbol detection especially under the receiver non-linearity. 
Connections to the boosting method may provide insights on designing the number of neurons in each layer. 
Another interesting direction for future work is how such RC-based detection methods can be combined with transmit-side precoding to jointly optimize the link performance with limited CSI feedback at the transmitter. 
The full potential of the RCNet symbol detection method is yet to be explored. 
From a theoretical standpoint, it would also be meaningful to analyze the functionality of each layer in the interference cancellation for a multi-user MIMO network. 

\appendix
\label{appendix_t_f_learning}
We now consider using alternative least squares to solve the problem (\ref{time_frequency_RC_weights_learning}). When ${\boldsymbol W}_{tout}$ is given, (\ref{time_frequency_RC_weights_learning}) can be rewritten as
\begin{equation*}
\label{time_frequency_RC_weights_learning2}
\begin{aligned}
    \min_{{\boldsymbol w}_{fout}(n)}& \quad \|{\boldsymbol {\tilde F}}(n){\boldsymbol S}{\boldsymbol W}_{tout}{\text {diag}}({\boldsymbol w}_{fout}(n))-{\boldsymbol Z}(n)\|_F^2 \\
    s.t. \quad &{\text{diag}}(|{\boldsymbol w}_{fout}(n)|) = {\boldsymbol I}, \quad\forall {n = 0,\cdots, N_{sc}-1}
\end{aligned}.
\end{equation*}
where ${\boldsymbol {\tilde F}}(n) \triangleq {\boldsymbol I}\otimes {\boldsymbol f}(n)$, ${\boldsymbol f}(n)$ is the $n$th row of a Fourier matrix, ${\boldsymbol Z}(n) \triangleq [{\boldsymbol z}_0(n)^T, \cdots, {\boldsymbol z}_Q(n)^T]^T$, and
\begin{align*}
{\boldsymbol S} \triangleq [{\boldsymbol S}_0&(N_{cp}:N_{sc}+N_{cp}-1,:)^T,\\
&\cdots,{\boldsymbol S}_{Q-1}(N_{cp}:N_{sc}+N_{cp}-1,:)^T]^{T}
\end{align*}
By spelling out the objective, the optimization problem is equivalent to 
\begin{equation*}
\begin{aligned}
    \min_{{\boldsymbol w}_{fout}(n)}& -{\text {Real}}({\text {Tr}}({\boldsymbol {\bar Z}}(n){\text {diag}}({\boldsymbol w}_{fout}(n)){\boldsymbol Z}(n)^H))\\
    s.t. \quad &{\text{diag}}(|{\boldsymbol w}_{fout}(n)|) = {\boldsymbol I}, \quad \forall {n = 0,\cdots, N_{sc}-1}
\end{aligned}
\end{equation*}
where ${\boldsymbol {\bar Z}}(n) \triangleq {\boldsymbol {\tilde F}}(n){\boldsymbol S}{\boldsymbol W}_{tout}$. In addition, 
\begin{align*}
{\text {Tr}}({\boldsymbol {\bar Z}}(n)&{\text {diag}}({\boldsymbol w}_{fout}(n)){\boldsymbol Z}(n)^H) \\
&= \sum_j { w}_{fout,j}(n){\text{Tr}}({ \boldsymbol {\bar z}}_{j}(n){\boldsymbol z}_{j}(n)^H)
\end{align*}
where ${\boldsymbol {\bar z}}_{j}(n)$ is the $j$th column of ${\boldsymbol {\bar Z}}(n)$, ${\boldsymbol { z}}_{j}(n)$ is the $j$th column of ${\boldsymbol { Z}}(n)$ and ${ w}_{fout,j}(n)$ is the $j$ the entry of ${\boldsymbol w}_{fout}(n)$. In order to minimize the objective function value, we can select
\begin{align}
\label{frequency_domain_W}
    \angle { w}_{fout,j}(n)
    = -\angle ({\boldsymbol z}_{j}(n)^H{ \boldsymbol {\bar z}}_{j}(n))
\end{align}
where $\angle$ represents the angle of a complex number. When $\{{\boldsymbol w}_{fout}(n)\}_{n=0}^{N_{sc}-1}$ is fixed in (\ref{time_frequency_RC_weights_learning}), ${\boldsymbol W}_{tout}$ is learned by
\begin{align}
\label{time_frequency_RC_weights_learning3}
    \min_{{\boldsymbol W}_{tout}}& \|{\boldsymbol {\tilde F}}{\boldsymbol S}{\boldsymbol W}_{tout}-{\boldsymbol {\hat Z}}\|_F^2 
\end{align}
where ${\boldsymbol {\tilde F}}\triangleq {\text{diag}}({\boldsymbol F}, \cdots, {\boldsymbol F}) \in {\mathbb C}^{QN_{sc}\times QN_{sc}}$ in which ${\boldsymbol F}$ is the Fourier matrix; 
\begin{equation*}
{\boldsymbol {\hat Z}} \triangleq [{\boldsymbol {\hat Z}}_0^T,\cdots,{\boldsymbol {\hat Z}}_{Q-1}^T]^{T},  
\end{equation*}
and
\begin{align}
{\boldsymbol {\hat Z}}_q \triangleq [{\boldsymbol {\hat z}}^T_q(0), {\boldsymbol {\hat z}}^T_q(1),\cdots ,{\boldsymbol z}_q^T(N_{sc}-1)]^T,
\end{align}
in which ${\boldsymbol {\hat z}}_q(n) \triangleq {\boldsymbol z}_q(n){\text{diag}}({\boldsymbol w}_{fout}^{*}(n))$. Therefore, the learning rule of ${\boldsymbol W}_{tout}$ is 
\begin{align}
\label{time_domain_W}
    {\boldsymbol W}_{tout} = {\boldsymbol S}^{+}{\boldsymbol {\tilde F}}^H{\boldsymbol {\hat Z}}
\end{align}

\bibliographystyle{IEEEtran}

\end{document}